\documentstyle[twoside,fleqn,espcrc2,epsf]{article}

\def\text#1{{\rm #1}}

\def\vek#1{\mbox{\protect\boldmath $#1$}}

\newcommand{\AmS}{{\protect\the\textfont2
  A\kern-.1667em\lower.5ex\hbox{M}\kern-.125emS}}

\def\BorderBox#1#2{%
\vbox{\hrule height #1\hbox{\vrule width #1%
	#2 \vrule width #1} \hrule height #1}}
\title{Renormalization of Currents for Massive Fermions%
\thanks{Presented at Lattice '98, 14--18 July 1998, Boulder, Colorado}
 \hfill\normalsize FERMILAB-CONF-98/332-T
}

\author{Andreas S. Kronfeld and Shoji Hashimoto%
	\address{Theoretical Physics Group, 
	Fermi National Accelerator Laboratory, 
	Batavia, Illinois, U.S.A.}%
 	\hfill {\tt hep-lat/9810042}
}

\begin{document}

\begin{abstract}
The renormalization of vector and axial-vector currents for massive
fermions (in the ``Fermilab formalism'') is discussed.
We give results for non-degenerate masses, which are needed for
semi-leptonic form factors.
\end{abstract}

% typeset front matter (including abstract)
\maketitle

\section{INTRODUCTION}
\label{sect:intro}
Like many groups, we and our collaborators are 
calculating form factors of semi-leptonic decays,
\begin{eqnarray}
	B\to  D l\nu~\cite{Has98}, & \; &
	B\to\pi l\nu~\cite{Rya98},	\label{B decay} \\
	D\to  K l\nu~\cite{Sim98}, & \; &
	D\to\pi l\nu~\cite{Sim98},	\label{D decay}
\end{eqnarray}
and similar decays to the vector mesons $D^*$, $K^*$, or $\rho$.
This paper discusses the renormalization---including effects of the 
nonzero quark masses---of the weak currents inducing the transitions.
The form factors are needed to an accuracy of a few per cent, so it 
would be ideal to devise a fully nonperturbative renormalization 
program.
We show here how certain ratios of transitions induced by vector and 
axial-vector currents take care of most, but not all, of the 
renormalization nonperturbatively.
We also give the residual one-loop radiative corrections to these 
ratios.
They are a few per cent.
One should expect, therefore, that the uncalculated higher-order 
corrections will not be needed for some time.

Our analysis differs from that of the {\sl Alpha\/}
Collaboration~\cite{Sin97} in its treatment of the quark masses.
First, in no decay may the quarks be considered degenerate.
Second, the masses of the charmed and bottom quarks are 
larger than the energy scale of chromodynamics, $\Lambda_{\rm QCD}$.
Consequently, chiral Ward indentities are not helpful for normalizing 
axial-vector currents.
Third, these masses are usually, in practice, not small compared to the 
lattice cutoff.
Therefore, to isolate the leading mass dependence into coefficient
functions (rather than matrix elements) we apply the mass-dependent
improvement program of Ref.~\cite{KKM97} to the clover action.
In particular, we keep nonzero quark masses in Feynman diagrams; for a 
transition such as $b\to u$ the up quark's mass can always be set to 
zero later.

At tree level currents suitable for massive fermions are given
by~\cite{KKM97,Kro95}
\begin{eqnarray}
	{\cal V}^{cb}_\mu & = & 
	Z_{V^{cb}}\bar{\Psi}^c \gamma_\mu\Psi^b	\label{V}  \\
	{\cal A}^{cb}_\mu & = & 
	Z_{A^{cb}}\bar{\Psi}^c \gamma_\mu\gamma_5\Psi^b	\label{A}
\end{eqnarray}
where, with $\kappa$ denoting the hopping parameter,
\begin{equation} \begin{array}{r@{\;=\;}l}
	\Psi^f & \sqrt{2\kappa^f}
	\left(1+ad^f_1\vek{\gamma}\cdot\vek{D}\right)\psi^f,    \\
	\bar{\Psi}^f & \sqrt{2\kappa^f} \left(\bar{\psi}^f-
	ad^f_1(\vek{D}\bar{\psi}^f)\cdot\vek{\gamma}\right),
\end{array} \label{psi} \end{equation}
and $\psi^f$ is the field of flavor~$f$ in the hopping-parameter form 
of the action.
In the following, we speak of charmed and bottom quarks, but $c$ and 
$b$ really stand for distinct quark flavors.

The renormalization factors~$Z$ depend on both masses in the current.
At tree level the mass dependence factorizes
\begin{equation}
	Z_{V^{cb}}^{[0]}=Z_{A^{cb}}^{[0]}=
	e^{M_1^c/2}e^{M_1^b/2},
	\label{Z[0]}
\end{equation}
where $M_1^f$ is the rest mass of flavor~$f$.
Beyond tree level $Z_V\neq Z_A$ and the mass dependence no longer 
factorizes.
The main aim of this paper is to obtain the full mass dependence of
$Z_V$ and $Z_A$ in one-loop perturbation theory.
Note that for degenerate quarks the vector current's absolute
normalization can be used to define $Z_{V^{ff}}$ nonperturbatively.

The coefficient $d_1^f$ depends on the mass of flavor~$f$.
Beyond tree level there is no reason to expect that a universal 
rotation \`a la Eq.~(\ref{psi}) improves all currents.
One strategy is to define $d_1^f$ by requiring that the 
equal-mass vector current be conserved.
For other currents and when the masses are not the same, further 
improvement is attained by adding higher-dimension terms---%
$\partial_j\bar{\Psi}i\sigma_{ij}\Psi$, $\bar{\Psi}D_i\Psi$, and
$\partial_i\bar{\Psi}\gamma_5\Psi$---%
to the right-hand sides of Eqs.~(\ref{V}) and~(\ref{A}), as is done
with very light quarks~\cite{Sin97} and in nonrelativistic
QCD~\cite{Mor98,Boy98}.

\section{VECTOR CURRENT}
\label{sect:V}
Most of the renormalization can be captured by writing
\begin{equation}
	Z_{V^{cb}} = Z^{1/2}_{V^{cc}}Z^{1/2}_{V^{bb}}
	R_{V^{cb}}.
	\label{RV}
\end{equation}
The equal-mass factors $Z_{V^{\bar{q}q}}$ are obtained 
nonperturbatively, leaving the expansion
\begin{equation}
	R_{V^{cb}}=
	1 + \sum_{l=1}^\infty g_0^{2l} R^{[l]}_{V^{cb}}.
	\label{RVPT}
\end{equation}
The one-loop term $R^{[1]}_{V^{cb}}$ is set by requiring
\begin{equation}
	\frac{\langle c,\xi'|{\cal V}_0^{cb}|b,\xi\rangle}%
	{\bar{u}_{c,\xi'}\gamma_0u_{b,\xi}} = 1 + g^2 r^{[1]},
	\label{renorm}
\end{equation}
to one loop, where
\begin{equation}
	r^{[1]} = \frac{3C_F}{16\pi^2}
	\left(\frac{m_b+m_c}{m_b-m_c}\ln\frac{m_b}{m_c}-2\right).
	\label{r}
\end{equation}

The only Feynman diagram needed to compute~$R^{[1]}_{V^{cb}}$ 
is the vertex diagram.
Everything associated with the legs---self energies and the factors
$e^{M^b_1/2}e^{M^c_1/2}$---drop out by construction.
From the vertex diagram with unequal masses one must subtract the 
average of the diagram with equal masses.
From this combination one also must subtract the corresponding 
combination of the continuum vertex diagram.
This gives~$-R^{[1]}_{V^{cb}}$.
The resulting loop integral is ultraviolet and infrared convergent.
(Infrared cancellation occurs point-by-point if the masses in 
continuum propagators are taken equal to the corresponding kinetic 
masses.)

Before showing results it is worthwhile to anticipate the outcome.
\begin{figure} %[b]
\BorderBox{0pt}{
	\epsfxsize=0.46875\textwidth
	\epsfbox{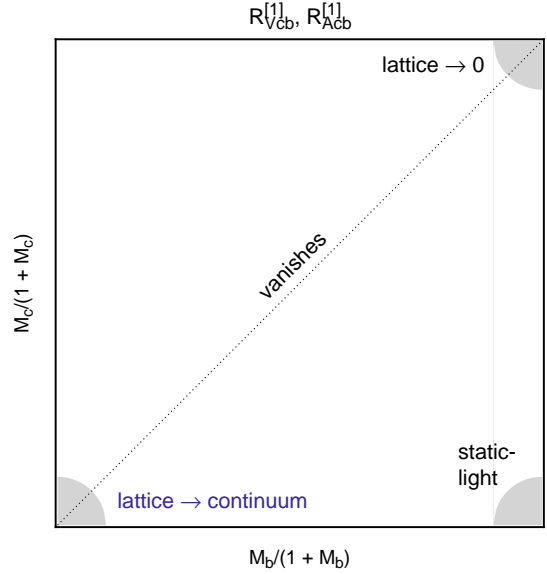}
}
\vspace*{-5mm}
	\caption[fig:sketch]{Limiting behaviors of
	$R^{[l]}_{V^{cb}}$, $R^{[l]}_{A^{cb}}$.}
	\label{fig:sketch}
\end{figure}
See Fig.~\ref{fig:sketch}.
$R_{V^{cb}}$ is symmetric under interchange of $b$ and $c$.
When the two masses are equal $R^{[l]}_{V^{cb}}=0$, $l\geq1$, by
construction.
When both masses are far below the lattice cutoff, 
$R^{[1]}_{V^{cb}}$ must vanish as $m_{b,c}a$ to a power.
(The linear term should be proportional to $1-c_{\rm SW}$, and it is.)
When both masses are much larger than the cutoff, the heavy-quark 
flavor symmetry (of the lattice action) ensures that the lattice 
contribution vanishes, i.e., $R^{[1]}_{V^{cb}}\to r^{[1]}$.
When one mass is larger than the cutoff and the other smaller,
$R^{[1]}_{V^{cb}}$ should resemble the radiative correction of 
a ``static-light'' current: dependence on the light mass should drop 
out, leaving a logarithm $(3C_F/16\pi^2)\ln M^b_2a$ plus a
constant~\cite{Ish98}.

\begin{figure} %[b]
\BorderBox{0pt}{
	\epsfxsize=0.46875\textwidth
	\epsfbox{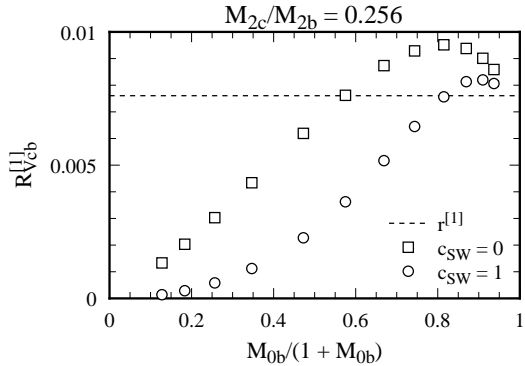}
}
\vspace*{-5mm}
	\caption{Radiative corrections to the $b\to c$ vector current.
	Circles (squares) denote $c_{\rm SW}=1$~(0).}
	\label{fig:RV}
\end{figure}

Fig.~\ref{fig:RV} bears out these expectations for a set of masses 
with $m_c=0.256m_b$, appropriate to charmed and bottom quarks.
These results have been obtained independently by the two authors.

To illustrate how to apply these results, let us consider the quark
masses relevant to our calculation of the semileptonic decay
$B\to Dl\nu$ at $\beta=5.7$ (and $\tilde{c}_{\rm SW}=1$),
for which $m_ba=3.9$ and $m_ca=1.0$~\cite{Has98}.
Reading off Fig.~\ref{fig:RV} one finds $R^{[1]}_{V^{cb}}=0.0076$.
For the strong coupling we consider the range $\alpha_s(\pi/a)=0.19$ to
$\alpha_s(1/a)=0.33$.
Thus, we multiply the bare matrix element with
\begin{equation}
R_{V^{cb}}= 1+0.0076\times 4\pi\times\alpha_s = 1.025(6).
\end{equation}
This estimate will be refined when we complete the calculation of the
BLM~\cite{BLM83} matching scale~$q^\star$.

\section{AXIAL-VECTOR CURRENT}
\label{sect:A}
\begin{figure} %[b]
\BorderBox{0pt}{
	\epsfxsize=0.46875\textwidth
	\epsfbox{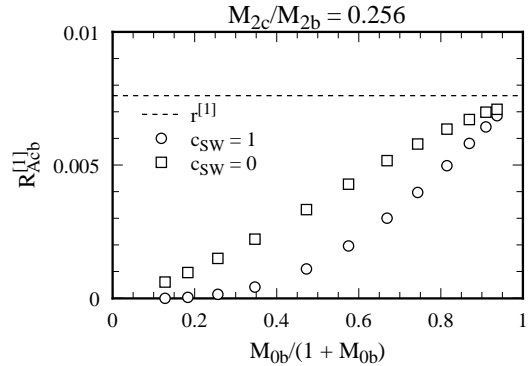}
}
\vspace*{-5mm}
	\caption[fig:RA]{Radiative corrections to the $b\to c$
	axial-vector current.
 	Symbols as in Fig.~\ref{fig:RV}.}
	\label{fig:RA}
\end{figure}

For the axial-vector current one could compute 
$Z_{A^{cb}}/Z_{V^{cb}}$, with the denominator from 
Eq.~(\ref{RV}).
For our analysis of $B\to D^*l\nu$, however, we need the ratio
\begin{equation}
	R_{A^{cb}}=\frac{Z_{A^{cb}}}%
	{Z^{1/2}_{A^{cc}}Z^{1/2}_{A^{bb}}}.
	\label{RA}
\end{equation}
The one-loop term $R^{[1]}_{A^{cb}}$ is set by analogy with 
Eq.~(\ref{renorm}), replacing ${\cal V}_0$ and $\gamma_0$ with 
${\cal A}_i$ and $\gamma_i\gamma_5$, but $r^{[1]}$ remains as in 
Eq.~(\ref{r}).
The limiting behaviors of~$R^{[1]}_{A^{cb}}$ are as
for~$R^{[1]}_{V^{cb}}$.
Our results are in Fig.~\ref{fig:RA}.

\section{OTHER COMPONENTS}
\label{sect:other}
Scattering matrix elements of ${\cal V}_i$ and ${\cal A}_0$ vanish as 
the three-momenta go to zero.
They can take the same renormalization factors as ${\cal V}_i$ and 
${\cal A}_0$, but commensurate effects are the one-loop corrections 
to~$d_1$ and to the coefficients of other higher-dimension
improvements.
The nonperturbative construction of a normalized, conserved
${\cal V}^{ff}_\mu$ implies that perturbation theory is needed only
for ``$m_b-m_c$'' and ``$V-A$''.

%In the equal-mass case, $d_1$ in the vector current can be computed 
%nonperturbatively, by requiring that the form factor multiplying 
%$\vek{q}=\vek{p}^b -\vek{p}^c$ vanish.
%\marginpar{make consistent}
%Perturbation is required, as in the renormalization factor itself, for 
%radiative corrections proportional to $m'-m$ and ``$V-A$''.

\section{SUMMARY}
\label{sect:summary}
By computing most of the renormalization nonperturbatively, it is 
possible to reduce the one-loop corrections.
This is achieved by requiring a correctly normalized, conserved 
vector current, or by extracting physics from a ratio of correlators, 
such that Eq.~(\ref{RA}) applies.
Indeed, the residual coefficients shown in Figs.~\ref{fig:RV} 
and~\ref{fig:RA} are small.
Hence, the uncalculated higher-order radiative corrections to 
semi-leptonic form factors are unlikely to be large.

\vspace*{3mm}
Fermilab is operated by Universities Research 
Association Inc., under contract with the U.S. Dept.\ of 
Energy.

\end{document}